\documentstyle[11pt,newpasp,twoside,epsf]{article}
\def\etal{{et\thinspace al.}\ }
\def\logg{$\log g$\thinspace}

\newcommand{\Teff}{\hbox{$T_{\rm eff}$}}

\newcommand{\ion}[2]{\mbox{#1\,{\small #2}}}

\def\md{log($\dot{M}$/M$_{\sun} {\rm yr}^{-1}$)\,}

\begin{document}
\title{FUSE observations of the central star of Abell 78}
 \author{Klaus Werner, Stefan Dreizler}
\affil{Institut f\"ur Astronomie und Astrophysik, Univ.\ T\"ubingen, Germany}
\author{Lars Koesterke}
\affil{NASA/GSFC, Greenbelt MD, U.S.A.}
\author{Jeff W. Kruk}
\affil{Department of Physics and Astronomy, JHU, Baltimore MD, U.S.A.}

\begin{abstract}
FUSE high resolution spectra of two PG1159 type central stars (K1-16 and
NGC\,7094) have revealed an unexpected iron deficiency of at least 1 or 2 dex
(Miksa \etal 2002). Here we present early results of FUSE spectroscopy of the
CSPN Abell~78. It is shown that iron is strongly deficient in this star, too.
\end{abstract}

The existence of hydrogen-deficient central stars (spectral types PG1159 and
[WC]) is probably due to a late thermal pulse. PG1159 central stars are very
hot (\Teff\ $>$100\,000\,K) and metals are highly ionized. The dominant
ionization stage of iron in the line formation region is \ion{Fe}{VII} and most
of its lines are located in the FUV region not accessible by HST. The lines are
expected to be rather weak and narrow. Due to the faintness of PG1159 stars,
iron abundances could not be determined until the advent of FUSE. It
was expected that the iron abundance would be essentially solar, because iron group
elements do not participate in any charged particle reactions occurring in low-
and intermediate mass stars. Therefore the detection of a strong iron
deficiency in two PG1159 type central stars was surprising (Miksa \etal 2002).

The central star of the planetary nebula Abell~78 is a very rare [WC]--PG1159
transition object showing spectral signatures of both early [WC] spectral type
(emission lines) and PG1159 type (absorption lines). The object was well
studied in the past, its atmospheric parameters were determined by NLTE model
atmosphere analyses (Koesterke \& Werner 1998):

\vspace{3mm}
\noindent
\Teff\,=110\,000\,K, \logg=5.5, \md= --7.3, {$v_\infty$}=3750\,km/s,

\vspace{2mm}
\noindent
He=33\%, C=50\%, N=2\%, O=15\% (by mass).
\vspace{3mm}

\noindent
A high resolution spectrum was taken with FUSE in order to look for iron
(Fig.\,1). No iron lines are detectable in the observed spectrum. Varying the
iron abundance in the NLTE models suggests that the absence of iron lines is
due to an underabundance of about two dex, compared to the solar value. For a
detailed discussion of possible explanations, e.g.\ neutron-capture
nucleosynthesis, see the respective papers by Herwig \etal and Werner \etal in
these proceedings.

\begin{figure}[!ht]
\epsfxsize=7.5cm
\epsffile{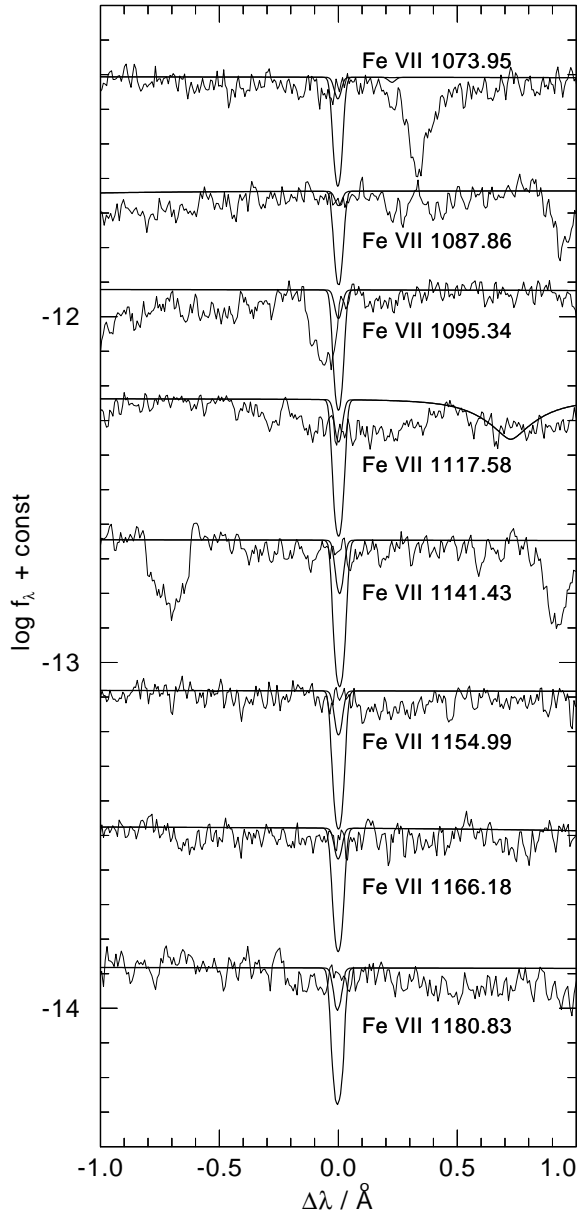}
\caption{
Details of FUSE spectrum of Abell~78 compared to theoretical line profiles,
centered around the strongest \ion{Fe}{VII} lines in the model. The model iron
abundances are 0.1 solar and 0.01 solar. No iron lines are detectable in the
FUSE spectrum. The apparent absorption line at 1095.2\AA\ is a detector artifact.
}
\end{figure}


\end{document}